\documentclass[conference]{IEEEtran}
\usepackage{cite}
\usepackage{graphicx}

\graphicspath{ {images/} }
\IEEEoverridecommandlockouts
\usepackage{cite}
\usepackage{amsmath,amssymb,amsfonts,bbm}
\usepackage{subfigure,bm,balance}
\usepackage[ruled,vlined]{algorithm2e}
\usepackage{graphicx}
\usepackage{textcomp}
\usepackage{xcolor}

\begin{document}


\title{QU-net++: Image Quality Detection Framework for Segmentation of Medical 3D Image Stacks}


\author{\IEEEauthorblockN{Sohini Roychowdhury}
\IEEEauthorblockA{Adjunct Faculty, Santa Clara University,\\
Email: roych@uw.edu}
\vspace{-0.5cm}
}

\maketitle
\begin{abstract}
Automated segmentation of pathological regions of interest aids medical image diagnostics and follow-up care. However, accurate pathological segmentations require high quality of annotated data that can be both cost and time intensive to generate. In this work, we propose an automated two-step method that detects a minimal image subset required to train segmentation models by evaluating the quality of medical images from 3D image stacks using a U-net++ model. These images that represent a lack of \textit{quality training} can then be annotated and used to fully train a U-net-based segmentation model. The proposed QU-net++ model detects this lack of quality training based on the disagreement in segmentations produced from the final two output layers. The proposed model isolates around 10\% of the slices per 3D image stack and can scale across imaging modalities to segment cysts in OCT images and ground glass opacity (GGO) in lung CT images with Dice scores in the range 0.56-0.72. Thus, the proposed method can be applied for cost effective multi-modal pathology segmentation tasks. 
\end{abstract}


\section{Introduction}
\label{intro}
Machine learning (ML) solutions for medical 3D image stacks rely on well annotated, high quality images to classify or detect regions of interest (ROIs) corresponding to pathology \cite{vnet}. The recent trend of re-using previously trained and deployed models and fine-tuning for specific use-cases, also known as \textit{transfer learning}, has significantly reduced the number of annotated image samples required to optimally train a ML model. However, there continues to be a need to identify a \textit{minimal training subset} of medical images for model fine-tuning purposes since the process of annotating medical images is both cost and time intensive. In this work, we present a novel two-stage system that identifies a minimal subset of images from medical 3D image stacks useful for training semantic segmentation models.

\begin{figure}[ht!]
    \centering
    \includegraphics[width=3.2in, height=1.5in]{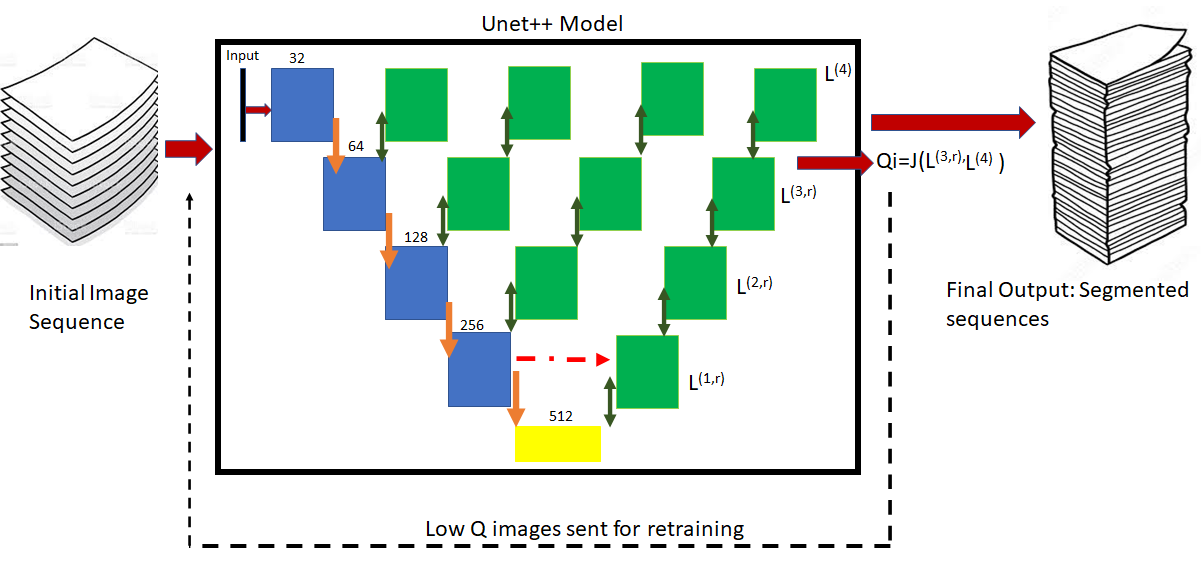}
    \caption{The proposed framework for minimal training data set detection. Quality training is detected using Jaccard score ($J$) between the final layers.}
    \label{system}
    \vspace{-0.5cm}
\end{figure}

Most existing works so far \cite{vnet} \cite{fewshot} rely on manual selection, random sampling, or previously trained ML models to identify batches of image data required to train a segmentation model. In this work, we present a novel framework shown in Fig. \ref{system} that scales across medical imaging modalities to identify a small training subset data. First, we identify an initial subset of the medical images/slices to be annotated based on the \textit{quality} of the images and the pixel variance captured within the annotated regions in images. Next, this image subset is used to train a 4-level multi-node U-Net++ model \cite{unetp}, such that the resized outputs from each of each level is analyzed, and if high variance is detected between outcomes of the final two layers, then the input image is considered to have new unlearned qualities/characteristics. Such an input image then gets appended to the training subset of images that need further annotation to fully train a segmentation model. Thus, at the end of the proposed two steps, a minimal training subset of images is identified to fully train a U-net++ model for semantic segmentation of all remaining images.

This paper makes two key contributions. First, we introduce a novel two-step image quality analysis method that starts from an initial set of 5-10 images to train/fine-tune a U-net++ model with deep supervision. The newly trained U-net++ model is then used to generate test segmentation masks for all remaining images from the 3D image stack. The resized test segmentation masks from the last two levels of the model are then used to detect images that represent a lack in quality training for the segmentation model. This process isolates 8-15\% of overall images from 3D image stacks to achieve state-of-the-art segmentation performances for pathology segmentation on test stacks. Second, we demonstrate the scope of transfer learning of the proposed U-Net++ based image quality detection model (QU-Net++) across medical image modalities and observe that kernel weights scale across computer tomography (CT) to optical coherence tomography (OCT) image stacks, thereby reducing the overall number of training samples needed for fine-tuning.
\section{Data and Methods}
\label{methods}
Descriptions of the Lung CT and OCT 3D image stacks and the proposed QU-net++ methods are presented below.

\subsection{Data: Lung-CT and OCT Image Stacks}
The first 3D image stack under analysis here is Lung CT images for COVID-19 segmentation of ground glass opacity (GGO) taken from the Kaggle dataset \cite{covid}. In this dataset, 100 individual images are annotated for GGO with available lung masks, as the \textit{Lung-CT-med} subset, and 829 images from a 3D volumetric scan are available as the \textit{Lung-med-rad} subset. Each image/slice is [512x512] in dimension and are resized to [256x256] for the U-net++ model. The second 3D image stack is that of OCT images form the OPTIMA cyst segmentation challenge (OCSC) dataset as described in \cite{fewshot}. We use 3 stacks of images per vendor-type along with annotations from observer $G_1$ from the Spectralis, Nidek, Topcon and Cirrus vendors. This results in 647 OCT slices from 3D image stacks that are cropped for the intra-retinal regions and resized to [256x256] for the U-net++.

\begin{figure}[ht!]
    \centering
    \includegraphics[width=3.0in, height=2.3in]{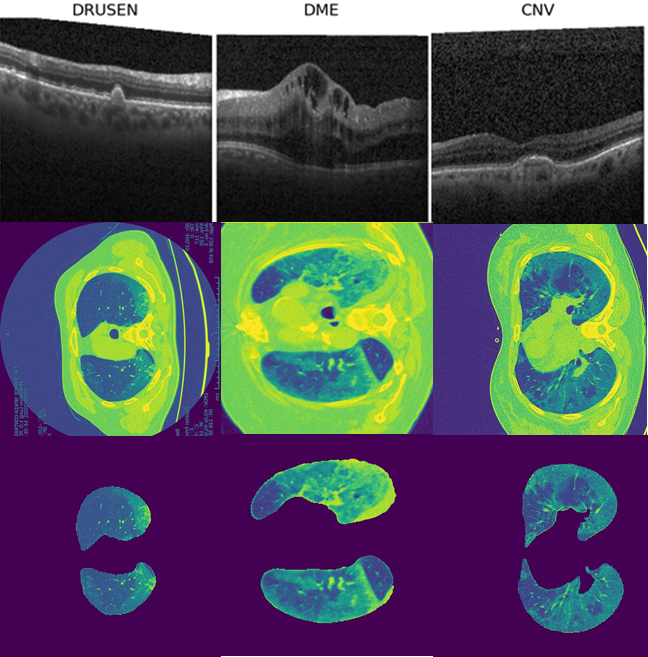}
    \caption{Samples of OCT and Lung CT images used in this work. Top row: cropped OCT images, Second row: Actual Lung CT images. Third row: Masked lung-CT images.}
    \label{fig:aug_image}
    \vspace{-0.5cm}
\end{figure}

Samples of the datasets used here are shown in Fig. \ref{fig:aug_image}. It is noteworthy that the grayscale OCT slices need to be cropped to include the intra-retinal layers as shown in \cite{fewshot}. Also, we observe that the CT slices may include metadata writing on them. Since the CT 3D image stacks includes masks to isolate the lung regions where the GGO regions exist, we utilize the masked-lung CT images as inputs for the segmentation, as shown in Fig. \ref{fig:aug_image}.
\subsection{Initial Training Image Set Selection}\label{initial}
\begin{figure}[ht!]
    \centering
    \includegraphics[width=3in, height=1.5in]{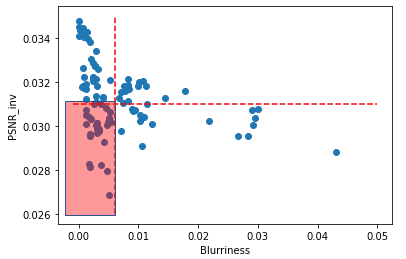}
    \caption{Example of initial training image set selection for the OCT stacks. Images in the highlighted bottom left quadrant are selected as $S_0$.}
    \label{sel}
    \vspace{-0.3cm}
\end{figure}

As a first step for minimal training subset identification, we begin with an unsupervised process of detecting an initial subset of images that represent \textit{good quality} of medical images with significant variations for the annotated regions. Here, the actual raw images are analyzed for \textit{blurriness} using the variance of Laplacian method \cite{blur}. \textit{Blurriness} is defined as the inverse of pixel variance upon applying the Laplacian operator on a grayscale image, such that a higher blurriness score indicates lower image focus and quality. Next, we evaluate the contrast in raw images using the inverse \textit{PSNR} metric, which represents the ratio between variance of pixels in a difference image, produced by subtracting a median filtering image from itself, over the maximum pixel strength in the image. A high value of this \textit{PSNR-inv} metric indicates high variance in pixel regions and low maximum foreground pixel strength, which indicates low contrast of the foreground regions. Thus, we isolate raw images that have less than average blurriness and \textit{PSNR-inv} metrics as an initial training image set $S_0$ to be used to train/fine-tune a U-net++ model, as shown in Fig. \ref{sel}. By varying the thresholds for blurriness and \textit{PSNR-inv} we can isolate about 10-20\% of original samples for initial training.

To further reduce the number of samples that require annotation from the initial set, we analyze the foreground region quality for each image in $S_0$. Here, we work with the masked images of the intra-retinal layers in OCT images and the masked Lung regions in the CT images. We evaluate two metrics for the pixels within the masked regions, namely: the coefficient of variation ($CoV$), defined as the ratio between variance and maximum pixel strength for all pixels within the ROI, and the mean pixel value within the ROI region.
Image samples that are within $\epsilon_0$ distance of others are considered to be \textit{similar} to the other samples and are thus eliminated from the initial training set ($S_0$). The goal is to minimize the initial set to less than 10 images per set that need annotation.

\subsection{QU-Net++ Model}
\begin{figure}[ht!]
    \centering
    \includegraphics[width=3.5in, keepaspectratio=True]{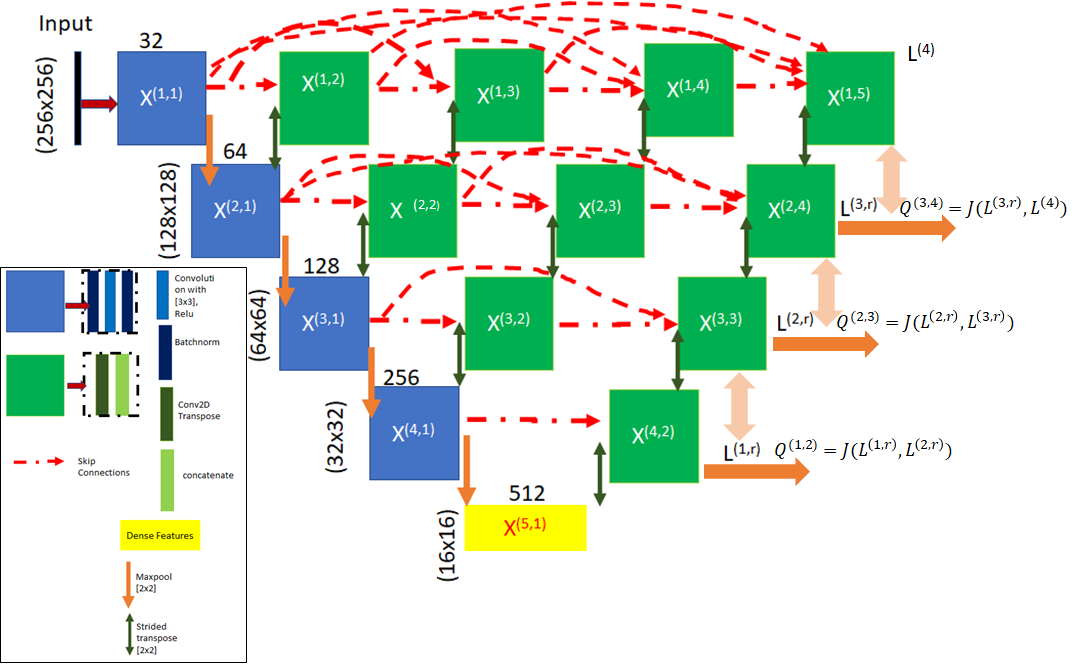}
    \caption{The QU-net++ architecture with quality evaluations ($Q$) between output layer masks at layers 1 through 4.}
    \label{model}
    \vspace{-0.5cm}
\end{figure}

Starting from the initial training image set, we train a U-net++ model \cite{unetp} and identify more images that represent a different image \textit{quality} than the image set previously selected for model training. Here, we apply a 4-level U-net++\cite{unetp} model to evaluate quality of the resized segmented masks, where the compositions of the encoder (convolution and pooling), decoder (transposed strided convolutions) layers and skip connections are shown in Fig. \ref{model}. For an optimal U-net++ model, we apply batch normalization to encoder layers only and dropout at layers $X^{(4,1)},X^{(5,1)}$ only \footnote{Github Code available at https://github.com/sohiniroych/QU-net-Plus-Plus}.

The primary difference between a U-net model and U-net++ \cite{unetp} model is the use of nested up-sampling layers and additional skip connections. For a U-net++ model the goal is to amplify signal strength at each transposed convolution layer (layers $X^{(4,2)},X^{(3,3)},X^{(2,4)}, X^{(1,5)}$) by concatenating with intermediate layers as shown in Fig. \ref{model}. This process increases the number of trainable parameters from 7,767,457 in a U-net to 9,045,540 parameters in the U-net++ model.   

For our application, we train a U-net++ model with a negative dice coefficient loss function shown in \eqref{dloss}, 
\begin{align}\label{dloss}
Loss=-\sum_{k=1}^{l_p}(\frac{2*[P(k)\cup Y(k)]}{[P(k)+Y(k)+1]}),
\end{align}
where, $l_p$ counts through all pixels in the segmented image, $P$ represents the predicted segmentation at level-4 ($L^4$) and $Y$ represents the annotated pathological ground-truth.

\begin{figure}[ht!]
    \centering
    \includegraphics[width=3.0in, height=2.7in]{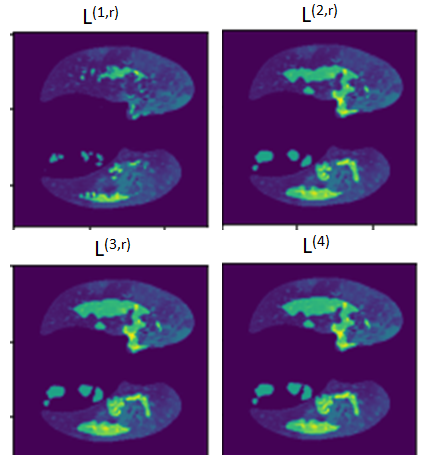}
    \caption{Examples of the 4 output levels from the U-net++ model for a sample Lung-CT-med image.}
    \label{sys}
    \vspace{-0.5cm}
\end{figure}

Next, we analyze the outputs at levels 1-4 ($L^{(1..4)}$) from the dense layer ($X^{(5,1)}$), using deep-supervising settings. The outputs at levels 1-3 are converted to the original image dimensions using the resizing ($^{r}$) operation. 
As the transposed convolutions move further away from the dense feature layer, only higher-order abstraction features at a global level get added to the semantic segmentation output. Thus, for a \textit{well-trained} U-net++ model, the initial transposed convolution layers closer to the $X^{(5,1)}$ layer bring major value to the semantic segmentation task while the farther away layers ($X^{(1,2)},X^{(1,3)},X^{(1,4)}$) have a lesser impact on the outcome. For this reason, we evaluate the intersection-over-union or Jaccard score ($J$) as representative of image \textit{content quality} ($Q_i$) for the resized level 3 and level 4 outcomes from the U-net++ model. If $Q_i$ for a particular test sample $i$ lies below a threshold $q_0$, then the image is considered to be important for model fine-tuning and added to the training set $S$. 
Examples of resized outputs from levels 1-4 for a Lung-CT-med image is shown in Fig. \ref{sys}. Here, the $Q_i$ score is 0.96 (high), from the outputs of levels 3 and 4. Thus, this image is not used for further model fine-tuning.

Algorithm \ref{algo} represents the steps for selecting the minimal training set of images $S_m$ needed to fine-tune a U-net++ model well. The input to this algorithm is the initial training image set ($S_0$) selected through raw image and annotation qualities described in Section \ref{initial}, and an empty set for $S_m$. The U-net++ model is run with deep-supervision to return the resized outputs at levels 1 through 4 and the quality index ($Q$) becomes a decisive factor if the image must be used for further fine-tuning of the U-net++ model or not. Once the minimal training set ($S_m$) is identified with $m$ samples, such that $m>=0$, the U-net++ model is further trained with these samples and the $L^{(4)}$ level output per test image is considered to be the final prediction per test image thereafter. Here, $I$ and $Y$ represent the raw image and the annotated segmented mask, respectively, and $n$ is the number of test images/slices.
\begin{algorithm}[ht!]
\SetAlgoLined
\KwOut{Minimal training dataset $S_m=\{I_m,Y_m\}$}
 \KwIn {Image sets: Initial training $S_0=\{I_0,Y_0\}$, Test: $\{I\}_{n}$, $S_m=\{\phi\}$}
 model$\longleftarrow$ U-net++$(S_0)$\;
 \For{$j=$ 1 \KwTo $n$}{
 $[L^{(1,r)},L^{(2,r)},L^{(3,r)},L^{(4)}]\longleftarrow$ model.predict($I_j$)\;
 $Q_j=\frac{L^{(3,r)}\cap L^{(4)}}{L^{(3,r)}\cup L^{(4)}}=J(L^{(3,r)}, L^{(4)})$\;
 \If{$Q_j < q_0$}{$S_m\longleftarrow S_m \cup I_j$}
  }
 \caption{Minimal Training Dataset detection}\label{algo}
 \vspace{-0.1cm}
\end{algorithm}
\section{Experiments and Results}\label{exp}
This work aims to optimally train a U-net++ segmentation model with the minimal number of training samples from 3D image stacks that can be identified based on image quality. To analyze the performance of the QU-net++ framework to isolate a minimal training set, we perform two experiments. First, we baseline the U-net and U-net++ models on the OCT and Lung CT stacks separately based on existing works in \cite{benchmark1} \cite{benchmark2} using randomly sampled training images. Second, we implement the proposed framework for minimal training set detection and analyze the segmentation performances of models trained on a fraction of images per stack on the remaining test images. The results and explanations are as follows.
\subsection{Baseline Segmentation: Random Sampling} 
Based on existing works in \cite{fewshot}, where 5-10 images per 3D image stack have been shown to train a U-net model for semantic segmentation, we randomly sample 25\% of the total number of slices per 3D image stack and use those images for training using U-net and U-net++ models. This process is repeated for 20 runs and averaged results are analyzed. The segmentation performances on the remaining test images are evaluated using the following metrics: precision ($Pr$) that represents the fraction of correctly predicted regions over all predicted regions; recall $Re$ that represents the fraction of correctly predicted regions over all actual ground-truth regions; Jaccard score ($J$) that represents intersection-over-union between the predicted and actual regions; Dice score ($D$) or negative of the loss function defined in \eqref{dloss}; accuracy ($Acc$) that represents the ratio of correctly classified foreground and background pixels over all pixels.
The numbers of training and test images are in first column and segmentation performances on the test images in comparison with existing works are shown in Table \ref{tab:num} .
\begin{table}[ht!]
\centering
\caption{Mean Segmentation Performances with varying training data.}
\scalebox{0.85}
    {
 \begin{tabular}{|c|c|c|c|c|c|c|}
 \hline
\#Train/\#Test&Model&$Pr$&$Re$&$J$&$D$&$Acc$\\ \hline
\multicolumn{7}{|c|}{Data: Lung-CT-med}\\\hline
Random (25/75)&U-net&0.59&0.51&0.36&0.49&0.95\\ 
Random (25/75)&U-net++&0.6&0.61&0.41&0.55&0.96\\ 
(698/117)&Fung et. al \cite{benchmark1}&0.38&0.82&0.32&0.43&-\\
{\bf (15/85)}&{\bf QU-net++}&{\bf 0.62}&{\bf 0.63}&{\bf 0.44}&{\bf 0.56}&{\bf 0.97}\\\hline
\multicolumn{7}{|c|}{Data: Lung-CT-rad}\\\hline
Random (207/621)&U-net&0.94&0.65&0.43&0.53&0.97\\
Random (207/621)&U-net++&0.98&0.63&0.54&0.62&0.98\\
(50/50)& FCN8 Fan et. al \cite{benchmark2}&0.91&0.53&-&0.471&-\\
{\bf(40/789)}&{\bf QU-net++}&{\bf0.99}&{\bf 0.63}&{\bf 0.65}&{\bf 0.64}&{\bf 0.99}\\\hline
\multicolumn{7}{|c|}{Data: OCT Stacks}\\\hline
Random (162/485)&U-net \cite{fewshot}&0.72&0.62&0.46&0.54&0.96\\
Random (162/485)&U-net++& 0.76&0.52&0.51&0.60&0.97\\ 
{\bf (67/580)}&{\bf QU-net++}&{\bf 0.86}&{\bf 0.65}&{\bf0.65}&{\bf 0.72}&{\bf 0.99}\\\hline
\end{tabular}
 }
\label{tab:num}
\vspace{-0.3cm}
\end{table}

\subsection{QU-net++ for Minimal Training Set Detection} 
\begin{figure}[ht!]
    \centering
    \includegraphics[width=3.3in, height=1.5in]{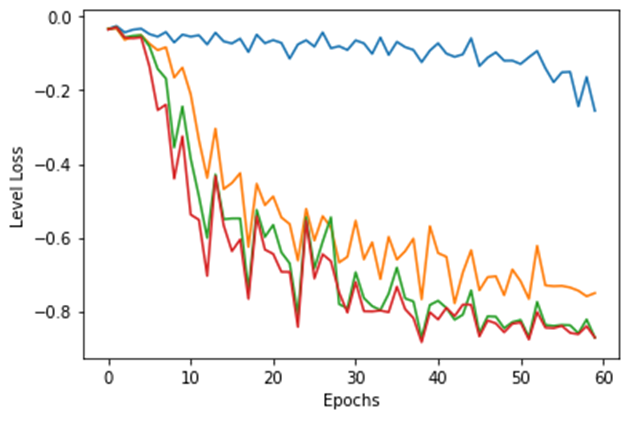}
    \caption{The loss curves at various levels for the OCT data set. Blue: $L^{(1,r)}$, Orange: $L^{(2,r)}$, Green: $L^{(3,r)}$, Red: $L^{(4)}$. }
    \label{loss}
    \vspace{-0.4cm}
\end{figure}

\begin{figure}[ht!]
    \centering
    \includegraphics[width=3.3in, height=2in]{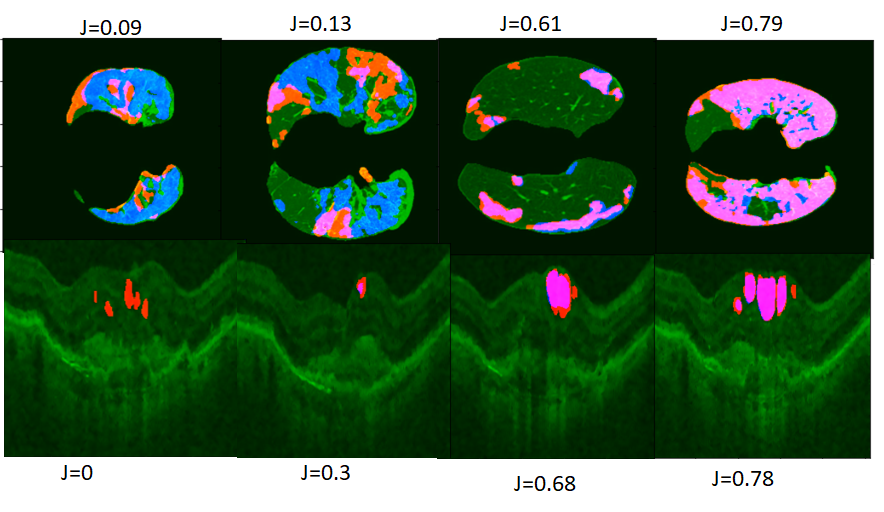}
    \caption{Examples of best and worst segmentation results at the end of fine-tuning U-net++ model with the minimal training data. Red regions: false negatives, Blue regions: false positives, Magenta regions: true positives.}
    \label{ex2}
    \vspace{-0.5cm}
\end{figure}

The proposed two-step framework is used for minimal training set detection and the performances of segmentation are shown in Table \ref{tab:num}. 
To fine-tune the U-net++ model, we apply data augmentations that include rotation, width, height and shear in range 0.2 with disabled vertical flipping. Adam optimizer with learning rate $10^{-4}$ is then used to train the U-Net++ model for 60 epochs with batch sizes of 20 images each.
Loss curves for training the U-net++ model with deep supervision on the minimal training set for the OCT stacks are shown in Fig. \ref{loss}. Here, we observe that upon training the loss curve trends for $L^{(3,r)}, L^{(4)}$ are very similar, further supporting our hypothesis for expecting similar outcomes from levels 3 and 4 from the U-net++ model.
Some examples of the finally trained model from the proposed framework are shown in Fig. \ref{ex2}. The first 2 columns represent GGO and cyst segmentations with low $J$ scores below 0.2. The last 2 columns represent segmentations with high $J$ scores above 0.6.

\section{Conclusion}\label {conclusion}
In this work we propose a novel image quality-based framework for isolating a minimal training set of images from 3D image stacks for semantic segmentation. We apply a trained U-net++ model with deep supervision and analyze the resized outputs from the final two levels to decide if the image under consideration should be used to further train the U-net++ model. This method extracts 8-15\% of all samples for training and results in overall pathology segmentation performances with Dice scores in the range 0.56-0.72. Future works will be directed towards extending this QU-net++ model to support multi-class segmentations.

\bibliographystyle{IEEEtran}
\bibliography{main}

\end{document}